\documentclass[final,1p,times,twocolumn]{elsarticle}
\usepackage{epsfig}
\usepackage{amssymb}
\usepackage{amsmath}
\usepackage{subcaption}
\usepackage{longtable}
\journal{Physics Letters A}
\usepackage{xcolor}

\DeclareMathAlphabet{\oldcal}{OMS}{cmsy}{m}{n}

\usepackage{hyperref}
\usepackage{multirow}
\usepackage{float}

\begin{document}

\title{The dynamic critical exponent $z$ for 2d and 3d Ising models from five-loop $\varepsilon$ expansion}

\author[label1,label2]{L.\,Ts.\,Adzhemyan}
\author[label1]{D.A.\,Evdokimov}
\author[label2,label3,label4]{M.\,Hnati\v{c}}
\author[label5]{E.~V. Ivanova}
\author[label1,label2]{\corref{cor1}M.\,V.\,Kompaniets}
\ead{m.kompaniets@spbu.ru} 
\author[label6]{\corref{cor2}A.\,Kudlis}
\ead{andrew.kudlis@metalab.ifmo.ru} 
\cortext[cor1]{}
\author[label1]{D.V.\,Zakharov}

\address[label1]{Saint Petersburg State University, 7/9 Universitetskaya Embankment, St. Petersburg, 199034 Russia}

\address[label2]{Bogoliubov Laboratory of Theoretical Physics, Joint Institute for Nuclear Research, 6 Joliot-Curie, Dubna, Moscow region, 141980, Russian Federation}

\address[label3]{Department of Theoretical Physics, SAS, Institute of Experimental Physics, Watsonova 47, 040 01 Ko\v{s}ice, Slovak Republic}

\address[label4]{Pavol Jozef \v{S}af\'{a}rik University in Ko\v{s}ice (UPJ\v{S}), \v{S}rob\'{a}rova 2, 041 80 Ko\v{s}ice, Slovak Republic}

\address[label5]{New Jersey Institute of Technology, 323 Dr Martin Luther King Jr Blvd, Newark, NJ 07102, USA}

\address[label6]{ITMO University, Kronverkskiy prospekt 49, Saint Petersburg 197101, Russia}

\begin{keyword}
renormalization group, critical dynamics, multi-loop calculation, critical dynamic exponent z, $\varepsilon$ expansion

\end{keyword}

\begin{abstract}
We calculate the dynamic critical exponent $z$ for 2d and 3d Ising universality classes by means of minimally subtracted five-loop $\varepsilon$ expansion obtained for the one-component model A. This breakthrough turns out to be possible through the successful adaptation of the Sector Decomposition technique to the problems of critical dynamics. The obtained fifth perturbative order accompanied by the use of advanced resummation techniques for asymptotic series allows us to find highly accurate numerical estimates of $z$: for two- and three-dimensional cases we obtain $\boldsymbol{2.14(2)}$ and $\boldsymbol{2.0235(8)}$ respectively. 
The numbers found are in good agreement with recent results obtained using different approaches.

\end{abstract}

\maketitle

\section{Introduction}

It is difficult to overstate the physical value of the concept of universality classes, proposed almost fifty years ago~\cite{green2017,nla.cat-vn4318275}. Universality  concerns a limited set of quantities including ratios of amplitudes and critical exponents~\cite{WILSON197475,RevModPhys.46.597,zjj1989,PELISSETTO2002549,V04}. The latter form a very canonical zoo. Static consideration of the critical behavior allows one to find all the exponents except one -- the critical dynamic exponent $z$. In the case of completely dissipative relaxational dynamics of the order parameter, such an exponent characterizes the critical slowing down linking the correlation length $\xi$ and the typical time of fluctuations $\tau$ as: $\tau\propto\xi^z$. Both quantities -- $\xi$ and $\tau$ -- diverge at criticality that is caused by the strong fluctuations in the system approaching the critical point. Numerical values obtained within various theoretical and experimental approaches for this exponent can roughly be estimated as $z\simeq 2$. Determination of the exact numerical deviation of $z$ from the value of two plays a crucial role in the theory.

In order to theoretically investigate this phenomenon one can address the one-component model A of critical dynamics, which is in fact the simplest dynamic generalization of the scalar $\phi^4$ field theory (see Ref.~\cite{HH77}) sufficient to describe the critical slowing down for systems with the nonconserved order parameter and energy.
The calculations of $z$ were carried out both using the Monte Carlo (MC) methods and different field theoretical (FT) approaches. 

Lattice calculations can be performed in different ways. The system can be quenched from the state above or below the transition directly to the critical point. A comprehensive review of different nonequilibrium universality classes for lattice systems is given in Ref.~\cite{Ordor2004}. For almost forty years, a large number of works have performed lattice calculations of $z$ for 2d and 3d Ising universality classes ~\cite{Wansleben1987,Wansleben1991,Mnkel1993,ito1993non3d,ito19932d,grassberger1995damage,Li1995,gropengiesser1995damage,Nightingale1996,stauffer1996flipping,Soares1997,Wang1997,Wang98,jaster1999short,godreche2000response,Ito2000,Nightingale2000,Lei2007,Murase2008,Murase2008,Collura_2010,hasenbusch2020}. 
Let us only pay attention to the recent work~\cite{hasenbusch2020}, where $z$ was estimated as $2.0245(15)$ in three dimensions. This result was obtained by considering the so-called improved Blume-Capel model that allows authors to eliminate the leading corrections to scaling.

As for FT calculations, there is also a vast variety of  results~(see Ref.~\cite{Folk_2006}). As is known, there are a number of different renormalization group (RG) approaches.

First, the RG analysis can be performed in fixed spatial dimensionality. The record-high result here is the four-loop calculations performed in Ref.~\cite{Prudnikov1997}. For 2d and 3d Ising universality classes for $z$ the authors have obtained $2.0842(39)$ and $2.0237(55)$, respectively, by using the advanced resummation strategies in Ref.~\cite{krinitsyn2006calculations}.

An alternative approach also applied to the problem is 
the so-called nonperturbative renormalization group (NPRG)~\cite{Canet_2007,PhysRevD.92.076001,DuclutDelamotte2017}. In Ref.~\cite{Canet_2007} the authors obtained the following estimates for two $2.16(1)$ and three $2.09(4)$ dimensions.
A noticeably different number from the latter was found in Ref.~\cite{PhysRevD.92.076001}: $2.025$. The authors of Ref.~\cite{DuclutDelamotte2017}, in turn, suggested a whole set of numerical estimates which correspond to application of different NPRG regulators (2d: $2.16$, $2.15$, $2.14$; 3d: $2.024$, $2.024$, $2.023$) and without them (2d: $2.28$; 3d: $2.032$).

The most canonical from the Wilson's time RG formalism, within which the present work is carried out, is the $\varepsilon$ expansion approach. The idea to shift the critical dimension by a small quantity ($\text{d}=4-\varepsilon$) played a crucial role. The two-loop result was found almost fifty years ago in Ref.~\cite{Halperin_1972}. Ten years later, a three-loop contribution was added, which has been record-high for a long time~\cite{Antonov1984}. The first attempt to calculate four-loop expansions was made in Ref.~\cite{ANS08} with subsequent resummation in Ref.~\cite{NSS9}. Only ten years later, however, the accuracy of four-loop calculations was notably improved~\cite{AIKV18} by means of a new \textit{diagram reduction technique}. Thanks to this technique, we are able to calculate the five-loop contribution to $z$. In such a high order of perturbation theory (PT), the reduction in the number of diagrams turns out to be crucial: from $\boldsymbol{1025}$ to $\boldsymbol{201}$. The calculation of the diagrams themselves requires special attention in the case of critical dynamics. The main difference is that the form of the integrand corresponding to these diagrams is much complicated due to time dependence. For this purpose, we address the Sector Decomposition (SD) technique~\cite{BINOTH2004375} as an effective method of accurate calculation of diagrams in the Feynman representation. It should be noted that initially SD was applied only to the static problems, where it has proven its efficiency. The authors of the present letter managed to adapt this method to critical dynamics diagrams~\cite{AIKV18}. The obtained five-loop expansion in combination with a novel resummation approach (we call it the \textit{free boundary condition method}, which was developed on the basis of the \textit{boundary condition method} suggested in Ref.~\cite{guida1998critical}) makes it possible to extract highly accurate estimates for $z$, which allows excellent consistency with the numbers that were found within the recent MC~\cite{hasenbusch2020} and NPRG~\cite{DuclutDelamotte2017} calculations. In this letter, we consider only some of the details of the calculation, giving special preference to the estimates of $z$.

Let us briefly cover the recent experimental results relevant to the problem. We note only that the experimental measurement of $z$ is fraught with enormous technical difficulties, which, so far, allows one to achieve only low numerical precision. Nevertheless, there is an extensive list of physical systems that are described by the one-component model A. For instance, in Ref.~\cite{NG15} the magnetoelectric dynamics in the multiferroic chiral antiferromagnet MnWO$_4$ was analyzed. The observation of critical slowing down of magnetoelectric fluctuations allowed the authors to make a conclusion about the validity of the theoretical description of this physical system within the model with overdamped magnetic 3d-Ising order parameter. The authors of other works~\cite{LF15,AuAgZn22018} studied the continuous phase transition of the AuAgZn$_2$ alloy by coherent x-ray scattering. Having quenched the alloy samples, they observed the motion of interfaces between ordered domains. Using the critical behaviour of the system, they also came to the conclusion that the critical dynamics of such a system corresponds to the one-component model A, although the number for dynamical critical exponent ($z=1.96(11)$) is in weak agreement with theoretical predictions. A result~($z\approx 2$) similar in accuracy  was obtained in Ref.~\cite{LB2002}.

The structure of the letter is arranged as follows. In Section~\ref{sec:method}, we briefly cover the model description and some features of RG approach within critical dynamics. In Section~\ref{sec:results}, we apply different resummation techniques in order to extract proper numerical estimates for $z$. Finally, we come to a conclusion in Section~\ref{sec:conc}.

\section{Method}\label{sec:method}
The action of the one-component model A is defined by a set of two scalar fields $\phi \equiv \{\psi,\,\psi'\}$ and can be expressed as follows~\cite{V04}:
\begin{equation}
S(\phi)
=\lambda \psi' \psi' + \psi'[ -\partial_t \psi + \lambda (\partial^2 \psi - m^2 \psi - \frac{1}{3!}g\psi^3)],
\end{equation}
where $\lambda$ is the Onsager coefficient and $g$ is the coupling constant. The model A is multiplicatively renormalizable. In this work, we use the minimal subtraction renormalization scheme (MS). The details of the renormalization procedure can be found in Ref.~\cite{AIKV18}. 
As a result, we obtain RG series for various observables. The important point here is the resummation strategies which should be applied to these expansions. Quite often, in addition to the direct application of various resummation techniques, the numerical estimates can be improved by various hints regarding the critical exponent behavior. For example, in Ref.~\cite{Bausch_1981} alternative expansion for $z$ in terms of the new parameter $\varepsilon'=d-1$ was obtained: $z=2+\varepsilon'+\varepsilon'^2/2+O(\varepsilon'^3)$. In addition, the authors resorted to information on the behavior of $z$ in $d=1,4$ spatial dimensionalities, as well as the behavior of the derivative. By combining this information and the canonical two-loop $\varepsilon$ expansion known at that time~\cite{Halperin_1972}, they constructed the following Pad\'e approximation:
\begin{equation}
    z-2=\frac{(d-1)(4-d)^2}{c_0+c_1d+c_2d^2},
\end{equation}
where the coefficients $c_i$ can be found from the condition of coincidence with the known $\varepsilon$ and $\varepsilon'$ expansions. The same trick was done in Ref.~\cite{hasenbusch2020} based on the known four-loop results~\cite{AIKV18}. 

In this letter, on the basis of the calculated five-loop contribution, we obtain numerical estimates for $z$ by means of the modified conformal mapping (CM) resummation technique that takes into account the strong coupling asymptotics. Pad\'e approximations are found both directly for five-loop $\varepsilon$ expansion and with the help of the known two-loop $\varepsilon'$ series.

\section{Results}\label{sec:results}
The five-loop $\varepsilon$ expansion of $z$ for the one-component model A reads:
\begin{eqnarray}\label{z}
	&&z=2+0.013446\varepsilon^2 + 0.011037\varepsilon^3-0.0055791(4)\varepsilon^4 + \boldsymbol{0.01774(31)} \varepsilon^5+O(\varepsilon^6).\,\,
\end{eqnarray}
The way of calculating the corresponding five-loop diagrams deserves a separate consideration; here we focus on extracting numerical estimates from the series~\eqref{z}.
\begin{figure} 
\center
\includegraphics[width=0.98\textwidth]{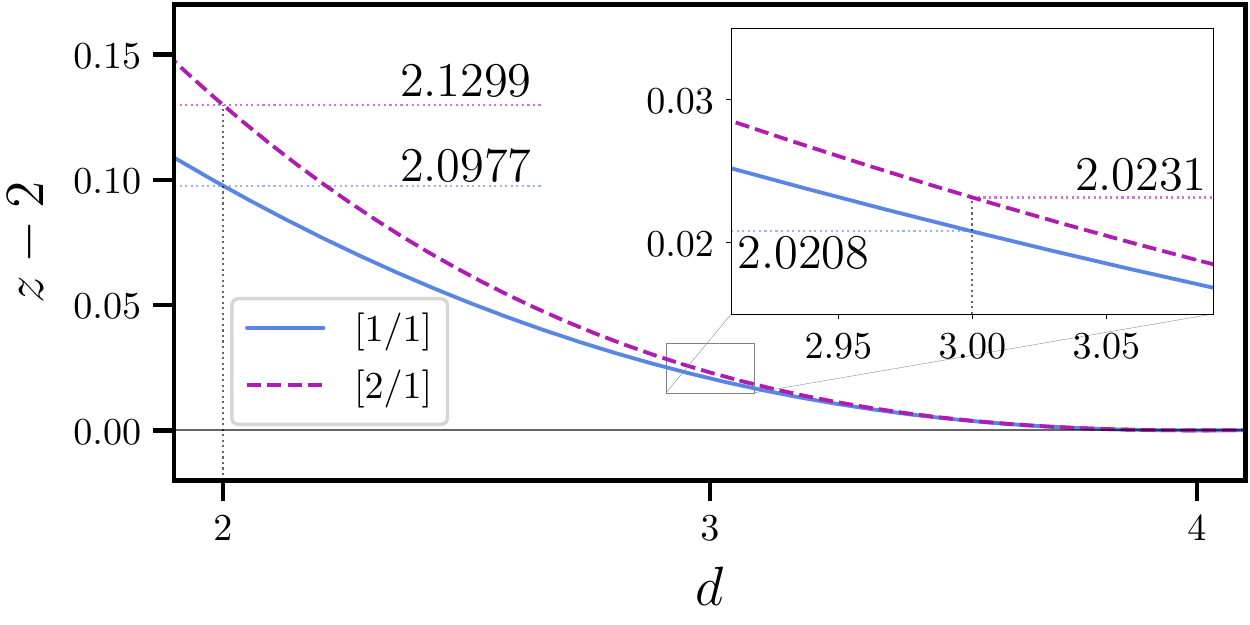} 
\caption{The dependence of Pad\'e approximants $[1/1]$ and $[2/1]$ on spatial dimensionality. The approximations are constructed based on $\varepsilon$ expansion~\eqref{z}. In the figure, we highlight the physically interesting values of $z$ given by both approximants.}
\label{pic:pade_simple}
\end{figure}

First, let us extract the desired numbers based on the Pad\'e approximations directly for the series~\eqref{z}. The structure of the corresponding approximants reads:
\begin{equation}\label{pade_direct} 
z-2 \approx (4-d)^2 \dfrac{P^{(L)}(d)}{Q^{(M)}(d)}=[L/M],\qquad L + M = 3 \, ,
\end{equation} 
where $P^{(L)}(d)$ and $Q^{(M)}(d)$ are polynomials in $d$ of order $L$ and $M$, respectively. Their coefficients are found from the condition that the truncated Taylor series of the given ratio $[L/M]$ (or specific Pad\'e approximant) coincides with the original part of asymptotic expansion. Eliminating approximations spoiled by dangerous poles, we are left with only two of them. Here, by spoiled approximants we mean those approximations, which have in the denominator the roots located in the physical domain in terms of spatial dimensionality: $d\in[1,4]$. The behaviour of these approximations as well as the corresponding estimates of $z$ for 2d and 3d cases are presented in Fig.~\ref{pic:pade_simple}.

\begin{figure}
\center
\includegraphics[width=0.98\textwidth]{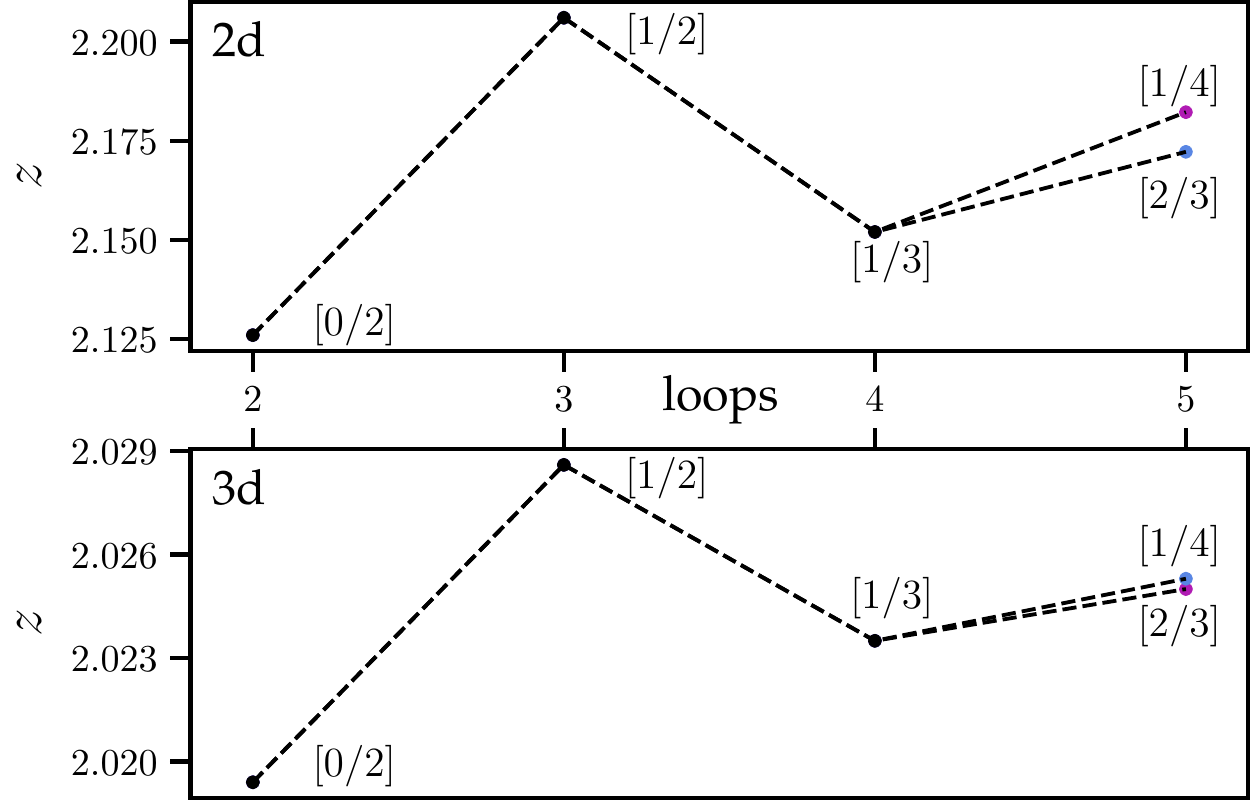} 
\caption{The behaviour of estimates for $z$ based on different Pad\'e approximations~\eqref{eq:pade_eps_pr} with regard to the number of loops. Except the five-loop case for each perturbation order only one approximant is not spoiled by poles.}
\label{pic:pade_eps_pr}
\end{figure}

Taking into account two-loop $\varepsilon'$ expansion~\cite{Bausch_1981} allows us to introduce two additional conditions on the coefficients for Pad\'e approximations:
\begin{equation}\label{eq:pade_eps_pr} 
z-2 \approx (4-d)^2 (d-1) \dfrac{P^{(L)}(d)}{Q^{(M)}(d)}=[L/M],\quad L + M=5\,. \end{equation} 
In order to demonstrate the convergence of estimates with increasing  PT order, adhering to this resummation strategy, we depicted the trends on the upper (2d) and lower (3d) panels in Fig.~\ref{pic:pade_eps_pr}. The corresponding five-loop estimates are $2.1772$ and $2.0251$, respectively. Despite a significant improvement in the numerical estimates, which is noticeable by their agreement with the results of other theoretical methods, these numbers still do not allow us to form a stable sample on the basis of which a final estimate could be proposed. 

\begin{figure}
\center
\includegraphics[width=0.98\textwidth]{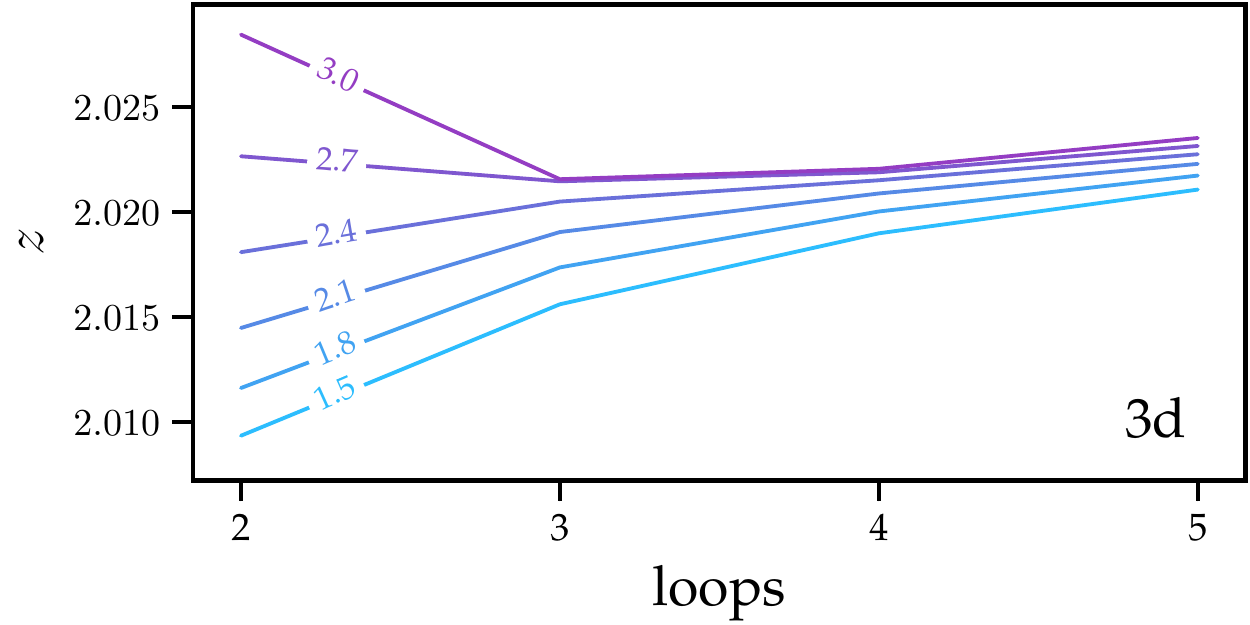} 
\caption{The behavior of estimates for critical exponent $z$ based on the order of PT for different values of $\lambda$. }
\label{pic:cb_wot_bc}
\end{figure}

 A more consistent way of resummation an asymptotic series $z-2=\sum_n A_n\varepsilon^n$ is to consider the Borel resummation technique, which is applicable to expansions with factorially increasing coefficients:~$A_{n}=C\, n!\,n^{b_{0}}(-a)^{n}(1+O(1/n))$, $n \to\infty$. For the one-component model A these parameters were calculated in Ref.~\cite{honkonen}: $a=1/3$ and $b_{0}=7/2$. As a result of the so-called Borel transformation, one can obtain the Borel image $F(x)$:
\begin{eqnarray}\label{eq:borel_im} 
&\sum\limits_{n} A_{n}\varepsilon^{n} = \int\limits_0^\infty dt \, e^{-t} t^b \sum\limits_{n} B_{n} (\varepsilon t)^n = \int\limits_0^\infty dt \, e^{-t} t^b F(\varepsilon t)\,. \,\, &
\end{eqnarray}
The analytical continuation of $F(x)$ has to be constructed, which is dictated by its finite  convergence  radius ($1/a$). One of the ways to obtain it is a conformal mapping of a particular form: $w(\varepsilon t)=(\sqrt{1+a\varepsilon t}-1)/(\sqrt{1+a\varepsilon t}+1)$. It transforms the real semiaxis into $[0,1)$ making the integration region within a circle of convergence.  Thus, it results in the following rewritten expansion:
\begin{equation}\label{eq:conf_borel} 
	\int\limits_0^\infty dt \, e^{-t} \, t^b \, \left(\dfrac{\varepsilon t}{w(\varepsilon t)}\right)^{\lambda}  \sum\limits_{n} B_{n}^{(w)} w^n\, , \quad b=b_0+\dfrac{3}{2},
\end{equation}
where $B_{n}^{(w)}$ can be calculated by means of reexpanding $F(\varepsilon t)$ in terms of the new variable $w$. The introduced multiplier -- $\left(\varepsilon t /w(\varepsilon t)\right)^{\lambda}$ -- is responsible for taking into account the strong coupling asymptotics: $A(\varepsilon)\sim\varepsilon^{\lambda}, \, \varepsilon\to\infty.$ As will be shown below, such manipulation also improves the convergence of estimates. However, the parameter of strong coupling asymptotics $\lambda$ for the model A is unknown. In Ref.~\cite{Kompaniets_2016}, the authors suggested to use $\lambda$ as a fitting parameter to improve the convergence. The criterion of its selection is the rate of convergence of the resummed results with sequential considering the known orders. The corresponding behaviour of 3d $z$ estimates for different $\lambda$ is presented in Fig.~\ref{pic:cb_wot_bc}. 
The observed tendency towards convergence of the results of calculations weakens with
increasing order of PT.

\begin{figure}
\center
\includegraphics[width=0.98\textwidth]{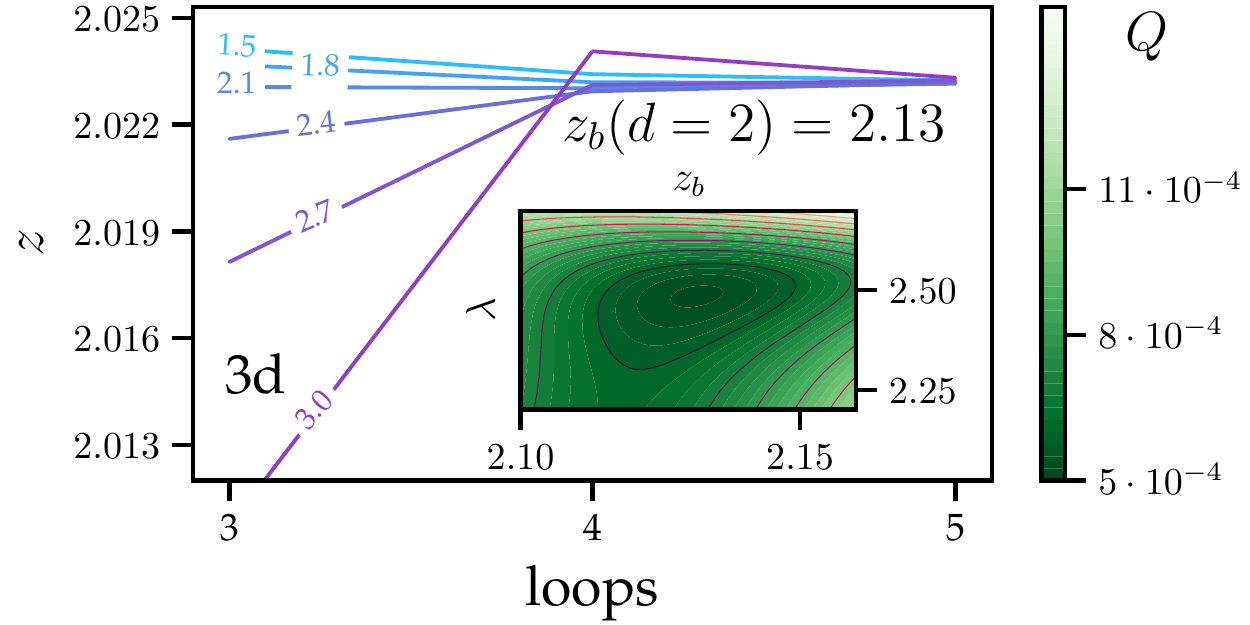} 
\caption{The behaviour of numerical estimates for critical exponent $z$ based on the order of PT for different values of $\lambda$ in the case of taking into account the boundary condition: $z_{2d}=2.13$.}
\label{pic:cb_with_bc}
\end{figure}

The situation can further be improved by resorting to the technique used earlier for calculating static exponents in Ref.~\cite{guida1998critical}. It was proposed to use the known values of critical exponents in other spatial dimensions. 
Unfortunately, for the dynamic critical exponent $z$ the exact value at any numerically helpful dimensions is unknown. 
However, we may use a trick according to which the value at a certain boundary point can be considered as a variable quantity determined only by the requirement of best convergence of the resummation procedure. The dependence of numerical estimates of $z$ in a three-dimensional case when the boundary value at two dimensions -- $z_b$ -- is assumed to be $2.13$ is presented in Fig.~\ref{pic:cb_with_bc}. As can be seen from the figure, the convergence has been noticeably improved in comparison with the case in Fig.~\ref{pic:cb_wot_bc}. As the optimal values of $\lambda$ and $z_b$, one should choose those that provide the best convergence to a given limit value based on a smaller number of terms of PT under consideration. This, in turn, corresponds to the fastest decrease in the slope of the straight line with an increasing number of loops. In order to formalize the convergence criterion, we write down the equation of the straight line passing through the points of four- ($z^{(4)}$) and five-loop ($z^{(5)}$) approximations as follows: 
\begin{eqnarray}\label{eq:slope_eq} 
z(l)=a(\lambda)(l-5)+b(\lambda)\,,
\end{eqnarray}
where $l$ is the number of loops,  $a(\lambda)=z^{(5)}-z^{(4)}$ and $b(\lambda)=z^{(5)}$. It is natural to demand the least sensitivity of the slope of the straight line $z(l)$ to the $\lambda$ variation that, in turn, compels the derivatives $\partial_{\lambda}a$, $\partial_{\lambda}b$, and $ \partial_{\lambda}^2b $ to take their minimum values. Therefore, as a criterion of convergence, one can choose the requirement of the minimum value of the following quantity:
\begin{equation}\label{eq:Q}
Q=\sqrt{\left(\partial_{\lambda}a\right)^{2}+\left(\partial_{\lambda}b\right)^{2}+\left(\partial_{\lambda}^2 b\right)^{2}}.
\end{equation}
We demonstrate the minimal value on the projection of the $Q(\lambda,z_b)$-surface as an inset in Fig.~\ref{pic:cb_with_bc}, which corresponds to the best convergence of estimates for the 3d case when the boundary point is $d=2$ and the calculation point is $d=3$. The choice of optimal parameters is not critical within the moderate range for $\varepsilon$.
To obtain the final estimates, averaging is carried out over both the boundary and calculation points. Our five-loop estimates obtained by means of this resummation technique are $2.13(4)$ and $2.023(2)$ for two and three dimensions, respectively. A comparison of the results obtained by means of the free boundary condition and without it shows that this modification of the CM resummation technique gives better agreement with the estimates of $z$ obtained in recent works~\cite{hasenbusch2020,DuclutDelamotte2017,PhysRevD.92.076001}.

In addition, we can address the way of numerical estimation of $z$ suggested in Ref.~\cite{Halperin_1972} via resummation of auxiliary quantity $c$, which is related to $z$ by $z=2+c\eta$, where $\eta$ is the static Fisher exponent. Taking into account the $\varepsilon$ expansion for $\eta$ from Ref.~\cite{BATKOVICH2016147,KP17} and the series~\eqref{z} for $c$, we obtain:
\begin{eqnarray}\label{c}
&&c=0.72609(1 - 0.18845\varepsilon + 0.22503(3)\varepsilon^2-0.379 (23)\varepsilon^3)+O(\varepsilon^4).\,\,\,\, \end{eqnarray}
Bearing in mind the fact that the coefficients of the series~\eqref{c} are small, it is hoped that the difficulties of analysis of the series manifest themselves only for the exponent $\eta$. In order to estimate $c$ from expansion~\eqref{c}, we find four- and five-loop Pad\'e approximants: $[1/1]$, $[2/1]$ and $[1/2]$. Averaging the obtained estimates for $c$ over two approximations and using the exact values $\eta=1/4$ for $d = 2$ and the most accurate $\eta=0.0362978(20)$ for $d=3$~\cite{Simmons_2017}, for the exponent $z$ we obtain $2.1512$ and $2.0236$ respectively.

Finally, we apply the KP17 method~\cite{KP17} which gives the most reliable results for $\phi^4$~\cite{KP17} and $\phi^3$~\cite{PhysRevD.103.116024, KOMPANIETS2021136331} models. This resummation technique includes three resummation parameters: $b$, $\lambda$, and $q$, where $b$ and $\lambda$ enter the expression~\eqref{eq:conf_borel}. The latter is responsible for the so-called homographic transformation, which allows one to take into account the potentially possible pole singularities of the $\varepsilon$ expansion. The selection strategy for these parameters is based on the their varying over a wide ranges in order to find the area of greatest stability of analyzed quantity. Here we have an additional source of error due to uncertainty of the numerical calculation of 4-th and 5-th terms, thus we add additional term related to this uncertainty. The corresponding results were added to Table~\ref{tab:results}.
\begin{table}[t!]
\center
\caption{Numerical estimates of $z$ obtained by means of different resummation strategies.}
\label{tab:results}
\setlength{\tabcolsep}{11.2pt}
\renewcommand{\arraystretch}{1.2}
\begin{tabular}{ccccccc}
\hline
\hline
Loops  & Dim. & Simple & Pad\'e via & Via coup.& Free& KP17\\ 
& & Pad\'e & $\varepsilon'$ exp. &  const. c  & b.c. & Ref.~\cite{KP17}  \\ 
\hline
\multirow{2}{*}{\large{4}} 
&2d&2.098 & 2.150 & 2.161 & 2.10(16) & 2.11(11)\\ 
&3d&2.021 & 2.024 & 2.024 & 2.021(5)& 2.023(4)\\
\hline
\multirow{2}{*}{\large{5}} 
&2d&2.130 & 2.177 & 2.151 & 2.13(4) &2.15(3)\\ 
&3d&2.023 & 2.025 & 2.0236 & 2.023(2)&2.0239(14)\\
\hline
\hline
\end{tabular}
\end{table}

\section{Conclusions}\label{sec:conc}
As a final estimate for the exponent $z$, we choose the weighted average calculated on the basis of all five-loop numbers from Table~\ref{tab:results}. The validity of this step is described in Ref.~\cite{PhysRevD.103.116024}. This tactic allows us to take into account an error bar of a particular sample element. Thus, for two- and three-dimensional cases we have $\boldsymbol{2.14(2)}$ and $\boldsymbol{2.0236(8)}$, respectively. These numbers are found in good agreement with the recently found results (2d: $2.1667(5)$ (MC~\cite{Nightingale2000}), $\sim 2.15$ (NPRG~\cite{DuclutDelamotte2017})) and (3d: $\sim 2.024$ (NPRG~\cite{DuclutDelamotte2017}), $2.0245(15)$ (MC~\cite{hasenbusch2020})) and can be considered as reliable ones.
Special attention, however, should be paid to the improvement of the results obtained by the CM resummation method together with the used strategy for determining the optimal values of the strong coupling parameter $\lambda$ and the optimal value at the boundary point $z_b$ by the convergence criterion $Q$. The applied method of the free boundary condition has noticeably accelerated the convergence of the resummation procedure, which may indicate its effectiveness in being applied to other models, where exact values of the quantities of interest in different spatial dimensions are unknown. Also, taking into account the $\varepsilon'=d-1$ expansion shows high efficiency in refining the results. This is especially pronounced when the four terms of the $\varepsilon$ expansion are taken into account, and additional information allows one to obtain a much more accurate value of $z$ compared to other methods. This is evidence in favor of using similar alternative expansions in the resummation procedure. Since the calculation of the six-loop coefficient of the $\varepsilon$ expansion for $z$ is a challenging computational problem, the results obtained may indicate the expediency of calculating the following orders in the $\varepsilon'$ expansion.

\section*{Acknowledgement}
We would like to thank R.Guida for the helpful discussion and sharing his notes.
The work of A.K. was supported by Grant of the Russian Science Foundation No 21-72-00108. The work of M.H. was supported by the Ministry of Education, Science, Research and Sport of the Slovak Republic(VEGA Grant No. 1/0535/21). We are grateful to the Joint Institute for Nuclear Research for allowing us to use their supercomputer “Govorun”.

\bibliographystyle{elsarticle-num}
\bibliography{Untitled}
\end{document}